\documentclass{article}
\begin{document}
\title{Behavioural Macroeconomic Policy: New perspectives on time inconsistency}
\author{Michelle Baddeley}

\maketitle

\begin{abstract}
This paper brings together divergent approaches to time inconsistency from macroeconomic policy and behavioural economics. Developing Barro and Gordon's models of rules, discretion and reputation, behavioural discount functions from behavioural microeconomics are embedded into Barro and Gordon's game-theoretic analysis of temptation versus enforcement to construct an encompassing model, nesting combinations of time consistent and time inconsistent preferences. The analysis presented in this paper shows that, with hyperbolic/quasi-hyperbolic discounting, the enforceable range of inflation targets is narrowed. This suggests limits to the effectiveness of monetary targets, under certain conditions.  The paper concludes with a discussion of monetary policy implications, explored specifically in the light of current macroeconomic policy debates.
\end{abstract}

\section{Introduction}
Contrasting analyses of time inconsistency are presented in macroeconomic policy and behavioural economics - the first from a rational expectations perspective \cite{KydlandPrescott77,barrogordon83a,barrogordon83b}, and the second building on insights from behavioural economics about bounded rationality and its implications in terms of present bias and hyperbolic/quasi-hyperbolic discounting \cite{Laibson,HarrisLaibson,Cohenetal,Fredericketal,ODonoghueRabin15}. Given the divergent assumptions about rational choice across these two approaches, it is not so surprising that there have been few attempts to reconcile these different analyses of time inconsistency.

This paper fills the gap by exploring the distinction between inter-personal time inconsistency (rational expectations models) and intra-personal time inconsistency (behavioural economic models). This raises questions about the role of institutional reform (e.g. inflation targetting and other monetary targets, central bank independence) versus behaviour change ("nudging") in ameliorating the negative impacts from these different forms of time inconsistency. In developing this analysis, sections I and II outline the key features of the rational expectations and behavioural economics models. Then, in Section III, the divergent approaches are reconciled via an encompassing model of strategic interactions between macroeconomic policy-makers and agents, building on Barro and Gordon's monetary policy model \cite{barrogordon83b}. Section IV outlines macroeconomic policy implications and conclusions.

\section{Time Inconsistency: Rational Expectations Models}
Mainstream macroeconomic theory assumes rational expectations and inter-temporal utility maximisation, building on microfoundations consistent with subjective expected utility theory \cite {NeumannMorgenstern53,Savage54}. Preferences - including time and risk preferences - are assumed to be independent, complete, stationary and consistent. Specifically for the rate of time preference, time-separable utility requires that preferences are consistent through time so that inter-temporal trade-offs are independent of when they occur. For example, with a stable rate of time preference, a choice between consumption in a year versus a year and a day would be equivalent to a choice between consumption in a decade versus a decade and a day. When the additional assumption of efficient financial markets is added, agents will balance their rate of time preference with the real interest rate. These assumptions are embedded in the "policy ineffectiveness" critiques of Lucas, Sargent, Wallace and others, which focussed on the limitations of discretionary demand management and its reliance on counter-cyclical monetary and fiscal policy levers  \cite{Lucas72,Lucas76,SargentWallace75,SargentWallace76}. 

Building on these insights, Kydland and Prescott proposed time inconsistency as an explanation for the stagflationary episodes of the 1970s \cite{KydlandPrescott77}.  Effectively operating as if in a strategic game with policy-makers, forward-looking rational agents will predict inflationary consequences from current expansionary policies, pre-empting inflationary erosion of their future real wages by increasing their nominal wage demands today. The consequences will be macroeconomic because, if all rational agents raise their nominal wage demands at the same time, then inflation will become a self-fulfilling prophecy as prices are pushed up by increasing wage costs. A sub-optimal Nash equilibrium will emerge in which rational agents' dominant strategy of increased nominal wage demands is neutralised by inflation, and everyone suffers from the inflationary bias which emerges as a consequence. 

Kydland and Prescott developed their concept of time inconsistency to capture the consequences of discretionary policy-making when rational private agents have opportunities to plan for the consequences of policies. Policies are time inconsistent when a policy devised to be optimal in one period will change the world (in this case by changing rational agents' expectations). Consequently, a policy designed to be optimal in the first period is no longer optimal in the next period, and the cause of its sub-optimality is itself i.e. its own impacts on private agents' expectations. In the case of discretionary demand management, private agents will shift their inflation expectations when discretionary policies are announced, rationally foreseeing the inflationary consequences of pushing an economy beyond full employment. So the policy designed to be optimal ex-ante, assuming static expectations, will not be optimal ex-post because rational agents’ expectations shift in response to the policy.  

What are the rational expectations alternatives to discretionary policy-making? Rational expectations theorists argued that, if the rules of the policy-making game are clear and transparent, then there will be no incentive for rational agents strategically to raise their wage demands. Sub-optimal outcomes emerge from strategic interactions between rational agents and macroeconomic policy-makers because policy-makers lack credibility in terms of their commitments to a full-employment target. If rational agents can be made to believe that policy-makers will stick to their full-employment targets, then the problem disappears. 

Important theoretical contributions to the analysis of credibility in macroeconomic policy-making include the models of Barro and Gordon, which add reputation-building by macroeconomic policy-makers into the rational expectations story \cite{barrogordon83a,barrogordon83b}. Barro and Gordon argue that concomitant high inflation and high unemployment were the consequence of non-credible commitments by monetary policy-makers, reflecting the fact that policymakers are tempted to push the economy beyond full employment in order to boost employment and production in the short term. When rational private agents know that policy-makers have these incentives to "cheat" by inflating the economy beyond full employment equilibrium, then those policies can be undermined. A non-cooperative equilibrium will emerge, in much the same way as described by Kydland and Prescott and, under these circumstances, the twin macroeconomic policy goals of controlling inflation and reducing unemployment to its natural rate cannot be attained. Nonetheless, policy-makers can be distracted from the temptation to cheat on their policy commitments in short-term if they are concerned by their reputations for policy-making credibility in the long-term. Far-sighted policy-makers will take into account rational private agents' perceptions of their credibility. If policy-makers are aware that their policy levers will be blunted in the future if their reputations are damaged today, then they will have incentives to change their strategies. There is an inter-temporal trade-off and so the policy-maker's rate of time preference becomes salient and the strategic policy game described by Kydland and Prescott shifts to a dynamic context and the Nash reversion strategy ("grim trigger" strategy) will be determined by discount rates. Specifically, forward-looking decision-makers, with a lower rate of time preference, will value their reputations more highly than decision-makers with a higher rate of time preference. It follows that forward-looking policy-makers will be less likely to cheat on their commitments, even though - if they were playing just a one-shot game - they would otherwise revert to a non-cooperative strategy \cite{Friedman71,AxelrodHamilton,Axelrod84}. 

In addressing these insights, received wisdom in macroeconomic policy-making circles shifted towards advocacy of inflation targetting and central bank independence as solutions. These insights had applications to other policy questions too, including industrial organisation and regulatory policy. For monetary policy, institutional reforms were widely adopted across the OECD, including in the UK and Australia - with New Zealand as a pioneer. Inevitably, however, debates around best practice have moved on, especially quickly in the case of monetary policy given the low inflation/low interest rate environment affecting most OECD economies in the aftermath of the 2007/8 financial crises. With the behavioural economics revolution, new insights can be added to these debates, as explored in the next section.

\section{Time Inconsistency in Behavioral Economics}
The rational expectations account has significant limitations - extensively explored in enduring debates between what are sometimes simplistically categorised as "Keynesians" and "Monetarists". These old debates reached an impasse in the 1970s, and contributed to significant ennui about macroeconomic theorising in policy-making circles, partly because there was little agreement about the behavioural assumptions.\footnote{These debates spawned a complex range of intermediate models - including New Keynesian and neo-Keynesian models, which combine traditionally Keynesian assumptions about sticky prices and imperfect markets, with the traditionally monetarist behavioural assumptions of optimisation and rational expectations. In the aftermath of the 2007/8 financial crises, for some macroeconomic policy-makers, the exclusion of the financial side from these models further eroded the credibility of mainstream macroeconomic theorising.} Alongside the seismic change in policy environment, economic theory generally has been shifting away from a focus on rational choice towards behavioural assumptions incorporating behavioural and psychological constraints. 

Broadly, behavioural economics allows that decisions are boundedly rational - not necessarily irrational, but limited by a range of cognitive constraints interacting with the informational constraints, including adverse selection and moral hazard, as applied extensively across other fields of economics too, e.g. by Akerlof and Shiller. Macroeconomics has been slow to embed these insights, partly because behavioural economics is largely confined to experimental data about individual decisions and choices. The simplifying assumptions of rational, independent, self-interested and homogenous agents that enable aggregation in conventional macroeconomic models are precluded in behavioural macroeconomic models, though empirical techniques such as agent-based modelling provide alternatives \cite{Baddeley16,Baddeley19}. Nonetheless, behavioural economics' reach is now extending as new macroeconomic insights are uncovered using the lens of behavioural economics. 

Temporal discounting is a promising area for behavioural macroeconomists generally, and specifically for macroeconomic policy where discounting assumptions can be adapted to build a deeper understanding of the nature and consequences of the strategic games between policy-makers and private agents, as explored in the preceding section. 

There are, however, some fundamental differences in definitions of time inconsistency that need to be addressed explicitly in order to build a coherent model. Specifically on the theme of time inconsistency - behavioural economics introduces substantive differences in its definitions/conceptions of time inconsistency relative to Kydland and Prescott's definition. The difference can be explained via a distinction between \textit{inter-personal time inconsistency} - consistent with the rational expectations account outlined above, and \textit{intra-personal time inconsistency}. Intra-personal time inconsistency is sometimes described by behavioural economists as the outcome of intra-personal struggles between different ‘selves’ \cite{Fredericketal}, building on Strotz's early insight that inter-temporal decision-making was characterised by “intertemporal tussles” between different incarnations of the individual through time, manifested when the preferences of the ‘present self” and future selves collide \cite{Strotz55}. These tussles manifest in present bias and preference reversals. For example, when a person is planning for a long distant future, they may believe themselves capable of resisting temptation, but when temptation becomes more immediate and tangible, then their preferences change and they are not able to resist temptation after all. Their preference for resisting temptation reverses. 

The behavioural concept of intra-personal time inconsistency connects with the more general literature on behavioural bias in behavioural economics.\footnote{Though, it should be acknowledged that there is considerable divergence across behavioural economists about the meaning and significance of  behavioural bias.} To summarise the different impacts of inter-personal time inconsistency versus intra-personal time inconsistency: in rational expectations models of inter-personal time inconsistency, rational agents' stable time preferences are characterised by an exponential discount function. Given dynamic strategic games between policy-makers and rational agents in the rational expectations models explored above, this creates what can be understood as a form of \textit{institutional} present bias.  Its behavioural corollary, emerging when  assumptions of perfectly rational decision-making and rational expectations are relaxed, is \textit{behavioural present bias}, and this is a product of intra-personal time inconsistency. The different impacts of behavioural present bias on inter-temporal decision-making are captured by embedding hyperbolic or quasi-hyperbolic discount functions into inter-temporal trade-offs instead of the standard exponential discount functions \cite{Laibson,HarrisLaibson}.\footnote{There is more conscilience in the definitions than commonly acknowledged because behavioural present bias is not precluded in seminal "orthodox" neo-classical accounts of inter-temporal decision-making. Early on, Samuelson acknowledged that exponential discounting was potentially an excessively restrictive assumption [needs ref].}

An empirical advantage for models incorporating behavioural discount functions is the wide-ranging experimental evidence from psychological and behavioural economic studies confirming the endemicity of present  bias in decision-making by humans and other animals, demonstrating that individuals' rate of time preference is not stable as assumed in standard economic discounted utility models \cite{Fredericketal,Cohenetal,ODonoghueRabin15}. Behavioural time inconsistency is manifested as shifts in an individual’s rate of time preference depending on the time horizons over which they are constructing their choices. This interpersonal time inconsistency problem creates a problem of ‘present bias’, i.e. a disproportionate focus on short-term rewards and a mismatch between long-run intentions and short-run actions. 

In the standard discounted utility model, the discount factor is assumed to be constant: 
\begin{equation}
max\sum_{t=1}^{\infty}u_\tau(c_\tau)D(\tau)d\tau
\end{equation}
where u is utility and c is consumption and $D(\cdot)$ is the discount function: 
\begin{equation}
D(t)=\delta^{t}=(\frac{1}{1+r})^t\approx e^{-rt}
\end{equation}
where $\delta$ is the discount factor and r is the discount rate. 

In behavioural dynamic models, the discount factor declines at a greater rate in the short run than in the long run and is inversely related to the length of delay in rewards. So the value of future rewards is disproportionately low relative to the value of current rewards.  Mathematically behavioural/intra-personal time-inconsistency can be captured using hyperbolic and quasi-hyperbolic discount functions, where the discount rate varies according to when the payoffs are received \cite{Laibson,HarrisLaibson}. The discount factor becomes: 
\begin{equation}
D(t)=\beta\delta^{t}=\beta{\big[\frac{1}{1+r}}]^{t} 
\end{equation}
where $\delta$ represents the time-consistent rate of time preference, and $\beta$ is the present bias parameter, capturing time-inconsistent preferences for immediate gratification. If $\beta=1$, then preferences are time-consistent; but if $\beta$ is less than 1 then the agent will over-weight short-term rewards relative to long-term rewards, and the inter-temporal Euler consumption relation will break down.

There will be heterogeneity in preferences and choices. Intra-personal time inconsistency may be moderated if agents embed pre-commitment mechanisms into their decision-making. Angeletos \textit{et al.} observe that some hyperbolic discounters value commitment, and thus hold illiquid assets as for them the cost of doing so is offset by the value of commitment \cite{angeletosetal}. O'Donoghue and Rabin postulate that different individuals will respond in different ways to the potential for pre-commitment, depending on their type, identifying four types of individuals \cite{ODonoghueRabin99,ODonoghueRabin01}. These types of agents are categorised according to two factors: first, the extent to which they are aware of their time-inconsistency; and second, what they do to overcome their predispositions towards present bias and preference reversals. Adapting these definitions to the macroeconomic policy-making ecosystem, policy makers differ in their reactions to time-inconsistency, and can be categorised as follows:

\begin{enumerate}
\item \textbf{Na\"{\i}ve}: These types are forward-looking but completely unaware of their time inconsistency and likelihood of preference reversal. They na\"{\i}vely assume that their future selves will behave tomorrow as they do  today; that their preferences formed in time t + n will be identical to those anticipated in time t.  Na\"{\i}fs do not take into account their own time inconsistency when planning future actions; they choose their plans as viewed from today's perspective. If $\hat\beta$ is the individual's own estimate of their quasi-hyperbolic present bias parameter, capturing their beliefs about their own potential for self-control, and $\beta$ is their actual present bias parameter, then - for na\"{\i}ve individuals $\hat\beta=1>\beta$.
In a macroeconomic policy context, na\"{\i}ve policy-makers would believe that they were making good policy choices for the long-term, and would be unaware when their policy choices are time inconsistent. This would increase the chances of sub-optimal Nash equilibrium outcomes, as predicted by the rational expectations models described above

\item 	\textbf{Resolute}: These individuals are aware of, and anticipate ex-ante, their own intra-personal time inconsistency and so bind themselves to pre-commitment strategies. For example, they could bind themselves with commitment mechanisms such as long-term contracts and illiquid investments.  If their pre-commitment strategies are effective in removing present bias, then $\hat\beta=\beta=1$. The "tie me to the mast" example of Ulysses is often used to illustrate this idea. The corollary for macroeconomic policy-makers is the policy-maker who binds themselves to a policy target, e.g. an inflation target, with costly sanctions for deviating from this target. Whilst these sanctions have been implemented, in theory, as part of central bank reform across the OECD from the 1990s onwards, sanctions on central bankers have been difficult to implement in practice.   

\item \textbf{Sophisticated}: These individuals backward induct to anticipate ex-ante their own dynamic time-inconsistency. Sophisticates are aware that their preferences change in the future and so decide not to participate, to avoid the negative consequences of inconsistency. The common analogy is Ulysses deciding to take a take a different route to avoid the “irresistible and deadly call of the Sirens” (Hey and Panaccione, 2011). For these types, $\hat\beta=\beta$ in theory, but without practical implications given their decision to avoid the inter-temporal conflict. In a macroeconomic policy context, this would be a policy-maker who anticipates the dilemma identified by Lucas and others, and therefore abstains from intervening to control the macroeconomy.

\item \textbf{Myopic}: Myopic types decide on the basis of static preferences. They are essentially uber-na\"{\i}fs in that they not only fail to recognise the pitfall of time inconsistency but also fail to recognise the dynamic nature of the problems they face. So they are not forward-looking at all: their present bias parameter approaches zero, leading to a situation equivalent to an infinite rate of time preference, and so their discount factor on future rewards approaches zero. A "one-shot game" discretionary policy-maker is myopic in this sense.\footnote{This could be interpreted as a link between short-termist policy and Keynes's famous quote about "in the long run we are all dead" as a justification for macroeconomic policy-making myopia. But Keynes continues: "Economists set themselves too easy, too useless a task, if in tempestuous seasons they can only tell us, that when the storm is long past, the ocean is flat again" (\textit{A Tract on Monetary Reform} (1923), p. 80). Keynes's words in their entirety could be more subtly interpreted as foreshadowing the policy-maker's time inconsistency problem: if we focus on the long-distant future then we will fail to recognise, and resolve, the temptations that policy-makers face in the short-term.} 

\end{enumerate}
O'Donoghue and Rabin (2001) also define a fifth type of person as partially na\"{\i}ve, i.e. individuals who are partially aware of their changing preferences but not entirely and so  underestimate their magnitude, that is: $\hat\beta\in(\beta,1)$ 

\section{A General Model: Behavioural-Strategic interactions}
In a macroeconomic policy framework, inflationary bias created by institutional present bias will be magnified in the context of behavioural present bias, but separating inter-personal and intra-personal time inconsistency to identify institutional present bias versus behavioural present bias is a complex task. This section explores some of the implications of these complexities specifically in the context of the macroeconomic policy debates, building on the Barro and Gordon model introduced above. 

\subsection{Barro and Gordon's reputational model of policy-making}
To recap: Kydland and Prescott describe a strategic game between rational private agents and policy-makers, explaining inflation bias as the outcome of the sub-optimal Nash equilibrium that emerges in the context of non-cooperative strategic decision-making \cite{KydlandPrescott77}. Building on Kydland and Prescott, Barro and Gordon develop the idea that sub-optimal Nash equibria emerge in the macroeconomy as the outcome of non-co-operative strategies between agents and policy-makers, but with additional complexity emerging from policy-makers' concerns about their credibility and reputation \cite{barrogordon83a,barrogordon83b}.  Barro and Gordon specify the policy-maker's objective function as:
\begin{equation}
z{\textsubscript{t}}=(a/2)(\pi{\textsubscript{t}})^2-b{\textsubscript{t}}(\pi{\textsubscript{t}}-\pi{\textsuperscript{e}}{\textsubscript{t}})
\end{equation}
where $a,b{\textsubscript{t}}>0$, $\pi{\textsubscript{t}}$ is inflation at time t, and  $\pi{\textsuperscript{e}}{\textsubscript{t}}$ are inflationary expectations at time t. To link with a non-accelerating rate of inflation target (NAIRU) target, this will be achieved when $\pi{\textsubscript{t}}=\pi{\textsuperscript{e}}{\textsubscript{t}}$.\footnote{Barro and Gordon do not refer directly to NAIRU and monetarism more generally focusses on a natural rate equilibrium (implying labour market clearing), but their analysis is also logically consistent with a NAIRU equilibrium concept (which allows for involuntary unemployment).} When the economy deviates away from the NAIRU, there will be a cost in terms of inflation bias, given by:
\begin{equation}
(a/2(\pi{\textsubscript{t}}){\textsuperscript{2}})
\end{equation}

But there will also be benefits from inflationary policies, e.g. from increases in employment and/or government revenue. These will be given by:

\begin{equation}
b{\textsubscript{t}}(\pi{\textsubscript{t}}-\pi{\textsuperscript{e}}{\textsubscript{t}})
\end{equation}

Taking into account these benefits and costs and expressing in expected present value terms, the policy-maker will be minimising the following loss function:
\begin{equation}
Z{\textsubscript{t}}=E[z{\textsubscript{t}}+(1/(1+r{\textsubscript{t}}))\cdot{z{\textsubscript{t+1}}}+(1/(1+r{\textsubscript{t}})(1+r{\textsubscript{t+1}}))\cdot{z{\textsubscript{t+2}}}+...]
\end{equation}
where $r \textsubscript{t}$ denotes the discount rate between period t and t+1, and the discount factor is given by:
\begin{equation}
q{\textsubscript{t}}=\frac{1}{(1+r{\textsubscript{t}})}
\end{equation}
In Barro and Gordon's model, when policy-makers transgress by cheating on their policy commitments to a NAIRU target, then they will be punished. This punishment is delivered via  enforcement from private agents, who adjust their expectations to the detriment of the policy-maker, making it harder for policy-makers to achieve their NAIRU targets in the future. With exponential discounting, the expected present value of this enforcement cost is given by:
\begin{equation}
Enforcement=E[q{\textsubscript{t}}(z{\textsubscript{t+1}}-z{\textsuperscript{*}}{\textsubscript{t+1}})]=\tilde{q}\cdot(1/2)(\bar{b}{\textsuperscript{2}})a
\end{equation}

The policy-maker balances this cost against their temptation, i.e. the benefits they will accrue in the short-term if they cheat on their commitments.
\begin{equation}
Temptation=(1/2)(\bar{b}){\textsuperscript{2}}/a
\end{equation}

So the policy-maker will cheat on their commitments where the benefits captured in the temptation relation (10) are greater than the costs captured in the enforcement relation (9).

\subsection{A Behavioural Reputational Model of Policy-making}
Note that the discount factor in Barro and Gordon's model is given by equation (8). But the trade-offs facing the policy-maker will change if behavioural insights about intra-personal time inconsistency are embedded into this game, specifically if the exponential discounting assumption is replaced with a behavioural discount function, e.g. Laibson's quasi-hyperbolic discounting function \cite{Laibson,HarrisLaibson}. Specifically, q can be replaced with the behavioural quasi-hyperbolic discount factor:

\begin{equation}
D(t)=\beta\delta^{t}=\beta[\frac{1}{1+r}]\textsuperscript{t} 
\end{equation}
where $0<\beta<1$

Temptation, is by definition, a short-termist strategy, so the Temptation relation is not changed by incorporating a behavioural discount function. The Enforcement relation, however, does change because it is about the present value of consequences in the future from cheating today. Allowing for present bias, this becomes:
\begin{equation}
Enforcement=\beta\delta\cdot(1/2)(\bar{b}{\textsuperscript{2}})a    
\end{equation}

Note that the present bias parameter is less than 1 i.e. $0<\beta<1$, so it follows that $\beta\delta<q$. Therefore, with the present bias associated with quasi-hyperbolic discounting, the present value of the enforcement costs from cheating are less, and therefore - \textit{ceteris paribus} - the likelihood that the policy-maker will cheat is increasing in the degree of present bias, i.e. as $\beta$ approaches zero. In this way, behavioural present bias magnifies the degree of inflationary bias seen in Barro and Gordon's baseline model. Note also that when $\beta=1$, then Barro and Gordon's rational expectations result holds because, in that case, $\beta\delta=q$.

\section{Policy Implications and Conclusions}

The model introduced in this paper takes basic insights about time inconsistency from the behavioural and rational expectations literature to capture dynamic strategic interactions between policymakers and private (rational) agents in the macroeconomy. Relaxing the assumption of rationality to embed present bias has implications in terms of the discounting of future consequences from current policy choices, leading to interactions between intra-temporal time-inconsistency and inter-temporal/institutional time-inconsistency. Thus this paper provides a behavioural economics alternative to the rational expectations account of strategic interactions between private agents and policy-makers, demonstrating how behavioural biases, specifically present bias in the context of time inconsistency, change policy-makers' trade-offs. Whilst applied here to macroeconomic policy-making, it also has implications for policy-making more generally, for example industrial regulation and public-sector infrastructure investment policy. 

What does it add to the policy debates? In the rational expectations literature, time inconsistency is essentially an inter-personal/institutional problem emerging from self-interested strategic interactions between rational optimising agents and policy-makers. Rational agents anticipate that policy-makers will have incentives to reverse initial policy announcements, reducing welfare relative to a position in which a policy-maker credibly commits to a stable policy. Aware of the possibility of policy reversal, individual agents will anticipate these policy reversals, and thus these policies will be ineffective. Overall, interactions between private agents and unreliable policy-makers deliver outcomes that are socially sub-optimal but, from an individual’s perspective - whether a private agent or a policy-maker, their decisions cannot be improved, which is why institutional reforms are important. In the context of macroeconomic policy, independent central banks are an example of an institutional reform designed to resolve self-interested conflicts between rational agents and policy-makers. Delegation to independent policy-makers, removed from political pressures, can help to reduce problems created by time-inconsistency, assuming that the delegated authorities are not prone to rent-seeking in other ways. 

This paper shows, however, that institutional present bias and time inconsistency are only part of the problem.  Problems are likely to be magnified in the context of behavioural present bias, when intra-personal time inconsistency magnifies inter-personal time inconsistency. In terms of solutions, well-designed institutions can play a role in mitigating time-inconsistency problems, if complemented by behavioural macroeconomic policy solutions. For example if effective pre-commitment mechanisms for policy-makers, e.g. long-term contracts, can be enforced, then this will ensure that policy-makers' decisions are more far-signted. There also connections with other areas of behavioural economics that develop new theoretical insights about trust and reciprocity. If trust can be built between private agents and policy-makers then that will operate to decrease the probability that either party will default on their announced commitments. 

In terms of future directions for this research, to assess the real world impacts of this analysis quantitative analysis is needed to infer something about the degree of present bias affecting policy-makers' decisions - more likely via calibrated simulations given the difficulties around identifying policy-makers' behavioural parameters from published statistics. In addition - theoretically, further analysis is being conducted to explore how the balance between temptation and enforcement will vary across the 4 types of agents identified by O'Donoghue and Rabin, with these insights applied to the policy-making eco-system.

\end{document}